\RequirePackage{lineno}
\documentclass[twocolumn,prl,aps]{revtex4}
\pdfoutput=1

\newcommand{\Npart}{\mbox{$N_{\mathrm{part}}$}}
\newcommand{\Ncoll}{\mbox{$N_{\mathrm{coll}}$}}

\newcommand{\energy}{\mbox{$\sqrt{s_{NN}}=2.76$~TeV}}

\newcommand{\sqn}{\mbox{$\sqrt{s_{NN}}$}}

\newcommand{\pp}{\mbox{p+p}}
\newcommand{\sumet}{\mbox{$\sum E_T$}}

\newcommand{\pt}{\mbox{$p_{\mathrm{T}}$}}
\newcommand{\et}{\mbox{$E_{\mathrm{T}}$}}

\newcommand{\zee}{\mbox{$Z\rightarrow e^{+}e^{-}$}}
\newcommand{\zmm}{\mbox{$Z\rightarrow\mu^{+}\mu^{-}$}}

\newcommand{\zll}{\mbox{$Z\rightarrow l^{+}l^{-}$}}
\newcommand{\ptZ}{\mbox{$p_{\mathrm{T}}^{Z}$}}

\newcommand{\yZ}{\mbox{$y^{Z}$}}

\newcommand{\mTAA}{\mbox{$\langle T_{\mathrm{AA}}\rangle$}}

\usepackage{atlasphysics}
\usepackage{slashed}
\usepackage{graphicx}
\usepackage{hyperref}

\usepackage{preprintcover}  
\PreprintCoverPaperTitle{Measurement of $Z$ boson Production in \PbPb\ Collisions at \energy\ with the ATLAS Detector}  
\PreprintIdNumber{CERN-PH-EP-2012-223}  
\PreprintCoverAbstract{The ATLAS experiment has observed 1995 $Z$ boson candidates in data corresponding to 0.15\,\inb\ of integrated luminosity obtained in the 2011 LHC \PbPb\ run at \sqn=2.76\,\TeV. The $Z$ bosons are reconstructed via di-electron and di-muon decay channels, with a background contamination of less than 3\%. Results from the two channels are consistent and are combined.  Within the statistical and systematic uncertainties, the per-event $Z$ boson yield is proportional to the number of binary collisions estimated by the Glauber model. The elliptic anisotropy of the azimuthal distribution of the $Z$ boson with respect to the event plane is found to be consistent with zero.}  
\PreprintJournalName{Phys. Rev. Lett.}  

\begin{document}


\title{Measurement of $Z$ boson Production in \PbPb\ Collisions at \energy\ with the ATLAS Detector}
\author{ATLAS Collaboration}

\begin{abstract}

The ATLAS experiment has observed 1995 $Z$ boson candidates in data corresponding to 0.15\,\inb\ of integrated luminosity obtained in the 2011 LHC \PbPb\ run at \sqn=2.76\,\TeV. The $Z$ bosons are reconstructed via di-electron and di-muon decay channels, with a background contamination of less than 3\%. Results from the two channels are consistent and are combined.  Within the statistical and systematic uncertainties, the per-event $Z$ boson yield is proportional to the number of binary collisions estimated by the Glauber model. The elliptic anisotropy of the azimuthal distribution of the $Z$ boson with respect to the event plane is found to be consistent with zero.\\
PACS numbers: 25.75.-q, 23.70.+j, 25.75.Cj, 25.75.Dw
\end{abstract}
\maketitle

Extensive studies of heavy ion (HI) collisions carried out by the experiments at the Relativistic Heavy Ion Collider (RHIC) at BNL, and the Large Hadron Collider (LHC) at CERN, have established that the hot and dense matter produced in HI collisions causes a significant energy loss on the energetic color-charge carriers propagating through such a medium~\cite{RHIC_review,LHC_review}. An understanding of this phenomenon requires measuring the unmodified production rates of the particles before they lose energy. The best candidates to perform such measurements are particles that do not interact via the strong force. 
The PHENIX experiment at RHIC measured the properties of photons~\cite{phenixphotons3}.  At the LHC, the CMS experiment reported results on photons and $W$ bosons~\cite{Chatrchyan2012256,CMS_W}.  The number of these bosons was found to scale with the number of incoherent nucelon-nucleon collisions.
Both the ATLAS and CMS collaborations have reported measurements of  \zmm\ production at  \sqn=2.76\,\TeV~\cite{Aad:2010aa,PhysRevLett.106.212301}, which show, within a limited statistical precision, the same scaling behavior.  This Letter presents a precise measurement of $Z$ boson production in \PbPb\ collisions at \sqn=2.76\,\TeV, using the di-electron and di-muon decay channels. The $Z$ boson production rate is measured as a function of  centrality, rapidity (\yZ),  transverse momentum (\ptZ), and orientation with respect to the event plane ~\cite{ATLAS-coord}.

The ATLAS detector~\cite{Aad:2008zzm} at the LHC covers nearly the entire solid angle around the collision point. It consists of an inner tracking detector surrounded by a thin superconducting solenoid, electromagnetic and hadronic calorimeters, and a muon spectrometer incorporating three superconducting toroid magnet systems. 
	
The inner-detector system (ID) is immersed in a 2~T axial magnetic field and provides charged particle tracking in the range $|\eta|<2.5$. The high-granularity silicon pixel detector covers the vertex region and is surrounded by the silicon microstrip tracker and the transition radiation tracker.

The calorimeters cover the range $|\eta|< 4.9$. Within the region $|\eta|< 3.2$, electromagnetic calorimetry is provided by barrel and end-cap high-granularity lead liquid-argon (LAr) calorimeters, with an additional thin LAr presampler covering $|\eta| < 1.8$.  The electromagnetic calorimeter is backed by a hadronic calorimter.  Forward calorimeters (FCal) are located in the range  $3.1<|\eta|<4.9$.

The muon spectrometer (MS) comprises separate trigger and high-precision tracking chambers that measure the deflection of muons in a magnetic field generated by superconducting air-core toroids. The precision chambers cover the region $|\eta| < 2.7$ with three layers of monitored drift tubes~(MDT), complemented by cathode strip chambers~(CSC) in the innermost layer of the forward region. The muon trigger system covers the range $|\eta| < 2.4$ with resistive plate chambers in the barrel, and thin gap chambers in the end-cap regions.

This analysis uses the 2011 LHC \PbPb\ collision data at \energy, obtained by the ATLAS experiment with integrated luminosity of approximately 0.15\,\inb.  The data sample for this study was collected using a three-level trigger system~\cite{Aad:2012xs}, which selected events with electron or muon candidates.

Electron candidates were identified at the first trigger level (L1) as a cluster of cells in the electromagnetic calorimeter, formed into $(\Delta\phi\times\Delta\eta)=0.1\times0.1$ trigger towers, within the range $|\eta|<2.5$, excluding the transition region between calorimeter sections ($1.37<|\eta|<1.52$). The cluster transverse energy was required to exceed $\et=14$\,\GeV.  

Muon candidates were selected using all three trigger levels. The L1 muon trigger searched for patterns of hits in the trigger chambers consistent with muons. If a muon had \pt\  exceeding 4\,\GeV, the event was accepted for further processing by the high-level trigger (HLT). The L1 muon algorithm also identified regions of interest~(RoI) within the detector to be investigated by the HLT.  In the HLT, the track parameters of each muon were recalculated by including the precision data from the MDT or CSC in the RoI defined by the previous trigger level. Muon candidates were reconstructed either solely from the MS or using combined data from the MS and ID. In addition to the events selected using the RoI-based muon trigger, the reconstruction was performed over the whole MS by the HLT to identify muons with $\pt>10$\,\GeV. The full scan searched all events in which a neutral particle signal was detected in each of two Zero Degree Calorimeters (ZDC) ($|\eta|>8.3$), or which contained an energy deposition in the calorimeters of $\et>10$\,\GeV.

In addition to the single-lepton trigger, each event had to pass the minimum-bias (MB) event selection, which required a timing signal coincidence of better than 3~ns between the MB trigger scintillators ($2.1<|\eta|< 3.8$), as well as the reconstruction of a collision vertex in the ID.  The total number of sampled events is 
$(1.03\pm0.02)\times10^{9}$~\cite{MB_confnote}.

Analyzed events are divided into centrality classes. Centrality reflects the overlap volume of the two colliding nuclei. Collisions with a small (large) impact parameter are referred to as central (peripheral).  The overlap volume is closely related to the average number of participant nucleons which scatter inelastically in each nuclear collision, \Npart, and to the average number of binary collisions between the nucleons of the colliding nuclei, \Ncoll.  Equivalently, \Ncoll\ may be defined as the average nuclear thickness function, \mTAA, multiplied by the total inelastic \pp\ cross section.    

The \PbPb\ collision centrality is measured using the scalar sum of transverse energy (\sumet) deposited in the FCal, calibrated at the electromagnetic energy scale~\cite{ATLAS:2011ag}.  The fraction of events with more than one \PbPb\  collision is estimated not to exceed 0.05\%, except for the most central 5\% of events in which the fraction does not exceed 0.5\%. A cut on the FCal energy of $\sumet<3.8$\,\TeV\ is applied to prevent contamination by events with multiple \PbPb\ interactions.  Glauber model calculations relate centrality to \Npart\ and \Ncoll, following the procedure documented in Ref. \cite{Aad2012330}. In the present sample, \Ncoll~(\Npart) ranges from $1683\pm130$ ($382\pm2$) for the most central class, 0--5\%, to $78\pm7$ ($46\pm3$) for the most peripheral class, 40--80\%.

The efficiencies of the electron and muon triggers are evaluated from $5.5\times10^{7}$ events selected with the MB trigger during the 2011 run. The MB  trigger required a transverse energy deposition of $\et>50$\,\GeV\ in the  calorimeters or a coincidence of both ZDC signals  and a track in the ID.  The average trigger efficiency for muons with $\pt>10$\,\GeV\ decreases from $98.2\pm 0.5$\% in peripheral events to $90.9\pm0.5$\% in central events, where the ID occupancy is higher. The average trigger efficiency for electrons with $|\eta|<2.5$ and $\et>20$\,\GeV\ is 
$98.1\pm 0.1\%$, independent of centrality.  The trigger efficiency for \Zmm\ decays ranges from $99.0\pm0.6\%$  in peripheral events to $95.0\pm0.9\%$ in central events.  For  \Zee\ decays the efficiency is $99.9\pm0.1\%$ independent of centrality.

For the \zee\ analysis, electron candidates are formed using the standard ATLAS reconstruction algorithm~\cite{Aad:2011mk}, requiring the matching of a track to an energy cluster in the electromagnetic calorimeter.    Electron selection is limited to $|\eta|<2.5$ and both electrons are required to have $\et>20$\,\GeV.  Following the  reconstruction requirements, further electron identification cuts are made to reject background.  The standard electron identification cuts~\cite{Aad:2011mk} used in the \pp\ environment are not suited to the \PbPb\ environment due to the large underlying event (UE) energy deposition in the calorimeter. To address this, a different set of cuts has been developed to accommodate the modification of the calorimeter variables by the presence of the UE. The cuts used are based on the energy balance between the track momentum and cluster energy ($E/p$), as well as calorimeter shower shape variables. Furthermore, the UE energy is estimated (following Ref. ~\cite{jet2}) and subtracted on an electron-by-electron basis to recover the proper electron energy.

The electron combined reconstruction and identification efficiency is evaluated in a Monte Carlo simulation using electrons from $7\times10^{5}$ {\sc Pythia}  (version 6.425) ~\cite{Sjostrand:2006za} \pp ~$\rightarrow$\ \zee\ events  with $66<m_{Z}<116$\,\GeV\ and $|\yZ|<2.5$ embedded into \PbPb\ events generated by the {\sc Hijing} event generator (version 1.38b) ~\cite{PhysRevD-44-3501}. The response of the ATLAS detector to the generated particles is modeled using GEANT4~\cite{Agostinelli2003250,ATLAS_sim}. The combined reconstruction and identification efficiency for electrons of $\et>20$\,\GeV\ ranges from 72\% to 76\% from central to peripheral events, with a common absolute uncertainty of 5.4\%.

For the \zee\ analysis, all electrons found in triggered events are paired with each other, requiring that at least one electron in the pair match a trigger object. The opposite-sign charged pairs with an invariant mass satisfying $66<m_{ee}<102$\,\GeV\ are accepted as signal $Z$ boson candidates.  The same-sign pairs in this window are taken as an estimate of the combinatorial background. In total, 772 opposite-sign pairs and 42 same-sign pairs are reconstructed.

In the \zmm\ analysis, single muons are reconstructed with several levels of quality~\cite{ATLAS-CONF-2010-036}. High quality muons are reconstructed in both the MS and ID with consistent angular measurements, as well as with a good match to the event vertex. At least one muon in each pair, matched to the trigger, is required to be of such quality. If the second muon in the pair has hit patterns in the MS and ID satisfying criteria of high reconstruction quality,  the minimum \pt\ threshold is set to 10\,\GeV\ for both muons. If the second muon fails this condition, both muons are required to satisfy $\pt>20$\,\GeV.

The muon combined reconstruction and identification efficiency is evaluated using muons from $5.3\times10^{5}$ {\sc Pythia} \pp ~$\rightarrow$\  \Zmm\ events with $66<m_{Z}<116$\,\GeV\ and $|\yZ|<2.5$ embedded into {\sc Hijing}  events. For muons with $\pt>20$\,\GeV, $|\eta|<2.5$ and associated to the event vertex, the reconstruction efficiency of the MS varies from $97\pm1$\% to  $98\pm 1$\% from central to peripheral events. Requiring a match between the MS and ID reduces the efficiency to $89\pm1$\% and $91\pm1$\%, respectively, due to track loss in the ID, predominantly at $|\eta|>1.5$. 

As in the \zee\ analysis, an invariant mass window of $66<m_{\mu\mu}<102$\,\GeV\ is used to define oppositely charged muon pairs as $Z$ boson candidates and same-sign charged pairs as a background estimate. In total, 1223 opposite-sign candidates and 14 same-sign pairs are reconstructed in the \zmm\ channel.

\begin{figure}[htb]
\begin{center}
\includegraphics[width=0.48\textwidth]{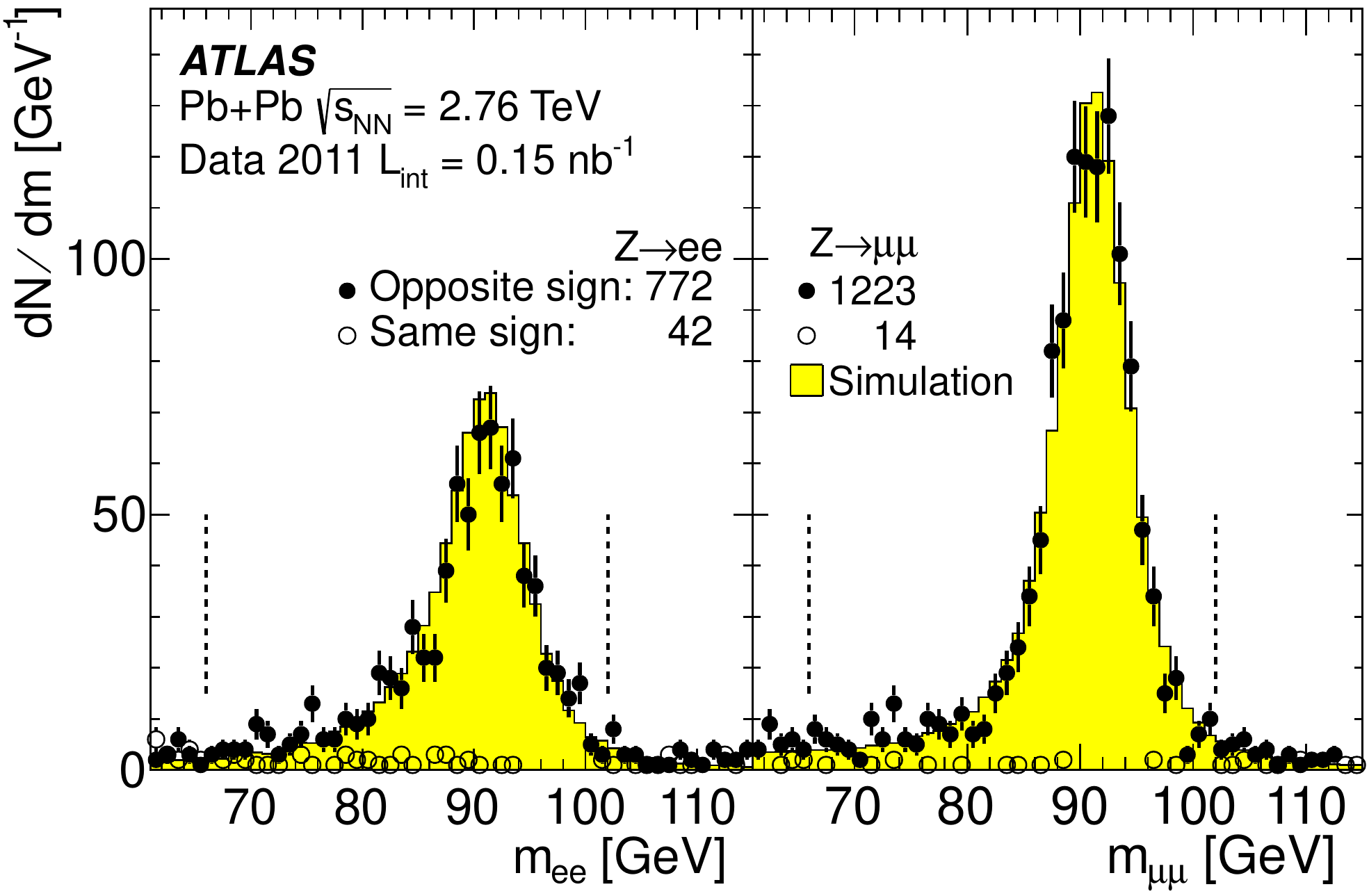}
\caption{The invariant mass distributions of \zee\ (left) and \zmm\ (right) candidates, integrated over momentum, rapidity, and centrality. Bars represent the statistical uncertainty.  The number of pairs with $66<m_{\ell\ell}<102$\,\GeV\ (marked by the vertical dashed lines) is listed.  The simulation is weighted to match the centrality distribution in data and normalized in the region $66<m_{\ell\ell}<102$\,\GeV.}
\label{fig:mass_peaks}
\end{center}
\end{figure}

The invariant mass distributions of the selected pairs together with estimated combinatorial backgrounds are shown in Fig.~\ref{fig:mass_peaks}, compared with the simulation normalized to the number of pairs in the region $66<m_{\ell\ell}<102$\,\GeV\ ($\ell=e,\mu$). In order to calculate the yield, the combinatorial background estimated with the same-sign pairs must be subtracted. Backgrounds from electroweak processes and top pair decays~\cite{Aad:2011dm} are small compared to the combinatorial backgrounds, and their contribution is accounted for in the systematic uncertainty related to the background.

The main sources of systematic uncertainty in both measurement channels are associated with the precision to which the corrections applied to the data can be calculated. In the \pp\ environment, the muon reconstruction efficiencies in data and simulation agree to 1\% (2\% for $\pt<15$\,\GeV)~\cite{ATLAS-CONF-2011-063}. The MS maintains low occupancy in the \PbPb\ environment. The difference in the fraction of muons reconstructed only in the MS, between data and simulation is used to estimate the systematic uncertainty on the reconstruction efficiency.  
To evaluate the uncertainty on the efficiency of the electron identification cuts, the efficiency is computed from the HI data using a tag-and-probe technique~\cite{Aad:2011mk} and compared to the efficiency computed from simulation. The systematic uncertainty due to momentum resolution is estimated by introducing additional momentum smearing to the simulation. The efficiency (resolution) uncertainties are $\approx$ 5.5\%  (2.5\%) for \zmm, and 8\% (2.5\%) in \zee; these estimates vary with \ptZ\ and \yZ. 

The trigger efficiency uncertainties are estimated by using alternative methods and comparing their results with those obtained from the MB data set.  For this comparison the simulation trigger efficiency is used, as well as the conditional trigger efficiency of a second lepton in a triggered pair reconstructed as a $Z$ boson.

For each \zll\ analysis, correction factors to account for the efficiency (relative to $Z$ bosons produced with $66<m_{Z}<116$\,\GeV) and detector resolution within the selected acceptance based on the simulation  are calculated differentially in event centrality, \ptZ, and \yZ.  In each decay channel, the correction factor is applied and the background, estimated by the same-sign pairs, is subtracted.  The two measurements are averaged with weights set by their respective uncertainties.   

\begin{figure}[htb]
\begin{center}
\includegraphics[width=0.49\textwidth]{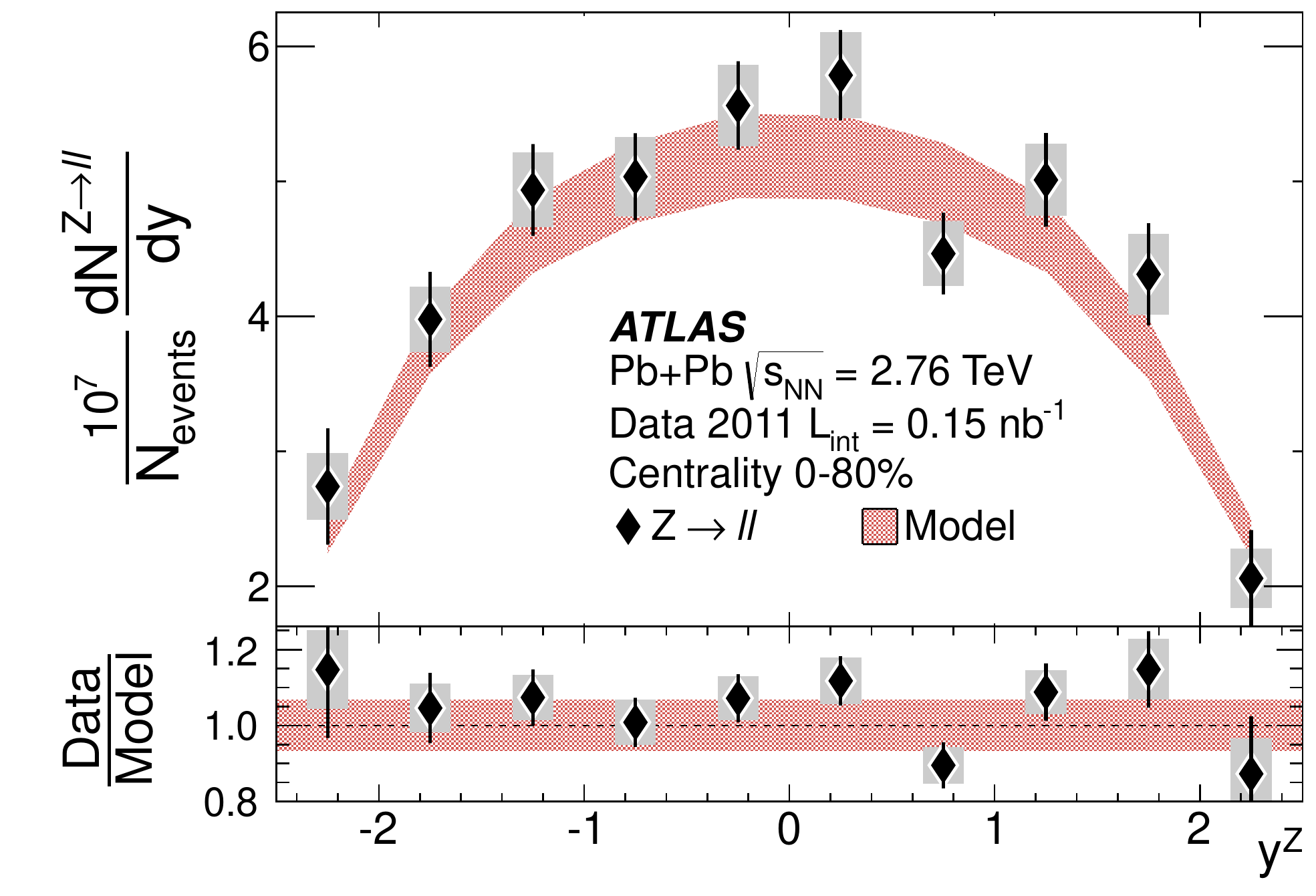}
\caption{The corrected per-event rapidity distribution of measured $Z$ bosons.  Bars and boxes represent statistical and systematic uncertainties, respectively.  The data are compared to the model distribution shown as a band whose width is the normalization uncertainty.} 
\label{fig:solo_rap}
\end{center}
\end{figure}

The fully corrected \yZ\ distribution is shown in Fig.~\ref{fig:solo_rap}. No centrality dependence of this shape is observed.  The data are compared to a model composed of {\sc Pythia} events normalized to the \zll\ cross section in \pp\ collisions at \energy\ taken from next-to-next-to-leading-order (NNLO) calculations used in Ref. ~\cite{ATLAS_pp_WZ} and scaled by \mTAA. Incorporating \pn\ and \nn\ collisions would increase the cross section by 3\%.  The shape is well reproduced by {\sc Pythia}, and the integrated yield is in good agreement with the \mTAA-scaled NNLO cross section. 

\begin{figure}[htb]
\begin{center}
\includegraphics[width=0.49\textwidth]{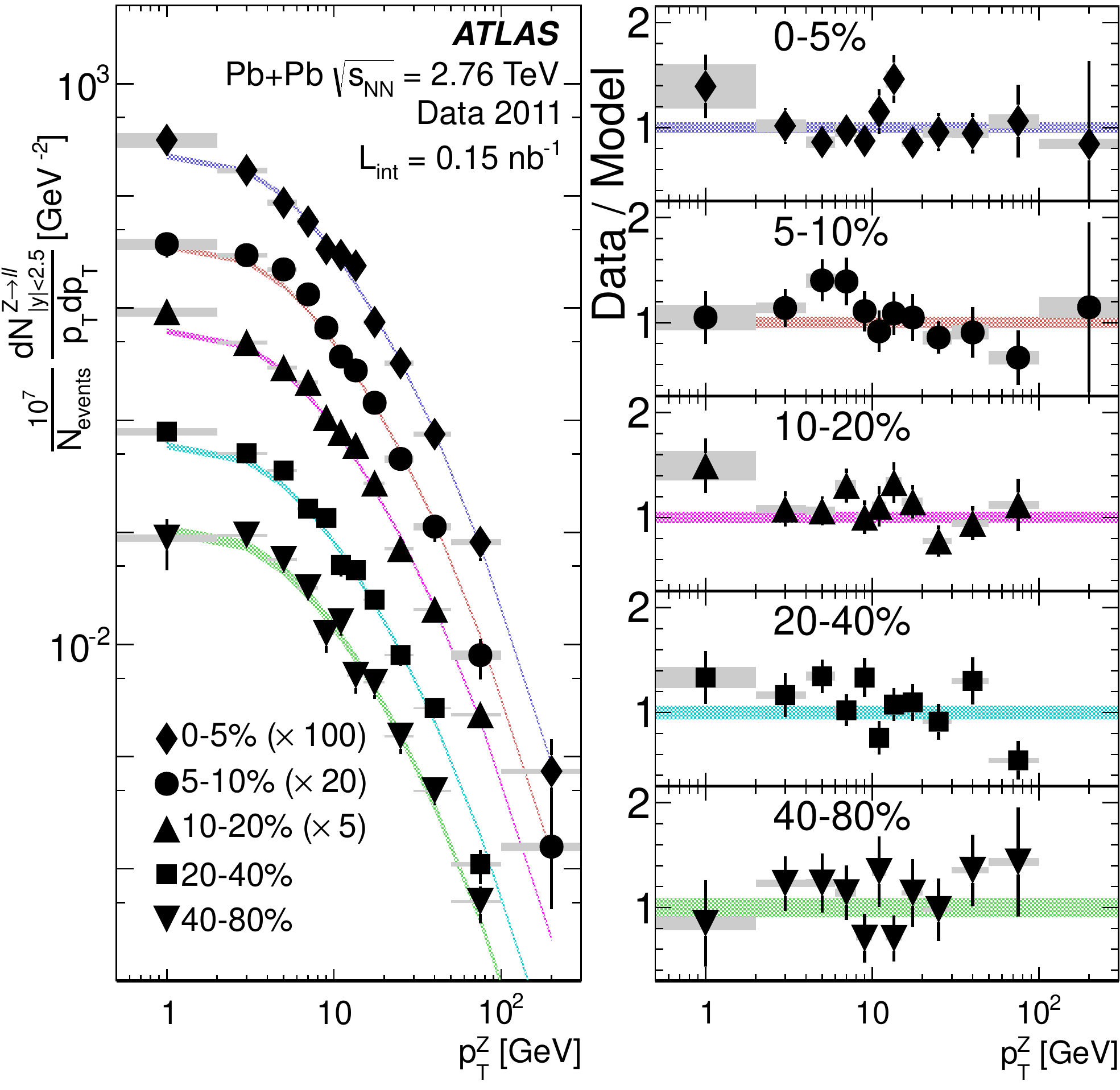}
\caption{Left: corrected per-event \ptZ\ spectra of measured $Z$ bosons in five centrality classes. The data are compared to a {\sc Pythia} simulation normalized to the NNLO \pp\ cross section and scaled by \mTAA, shown as bands.  Right: ratios of the data to the model in each centrality class.  Bars represent statistical uncertainties, boxes represent systematic uncertainties, and bands represent the normalization uncertainty.}
\label{fig:solo_pt}
\end{center}
\end{figure}

The fully corrected \pt\ distributions in five centrality classes are shown in the left panel of Fig.~\ref{fig:solo_pt} along with the model prediction. The shape as a function of \ptZ\ is well reproduced by {\sc Pythia}. The right panel of Fig.~\ref{fig:solo_pt} shows the ratios of the data to the {\sc Pythia} prediction scaled by \mTAA. The ratios are constant within uncertainties for all centrality classes over the range of measured \ptZ.

\begin{figure}[h!]
\begin{center}
\includegraphics[width=0.49\textwidth]{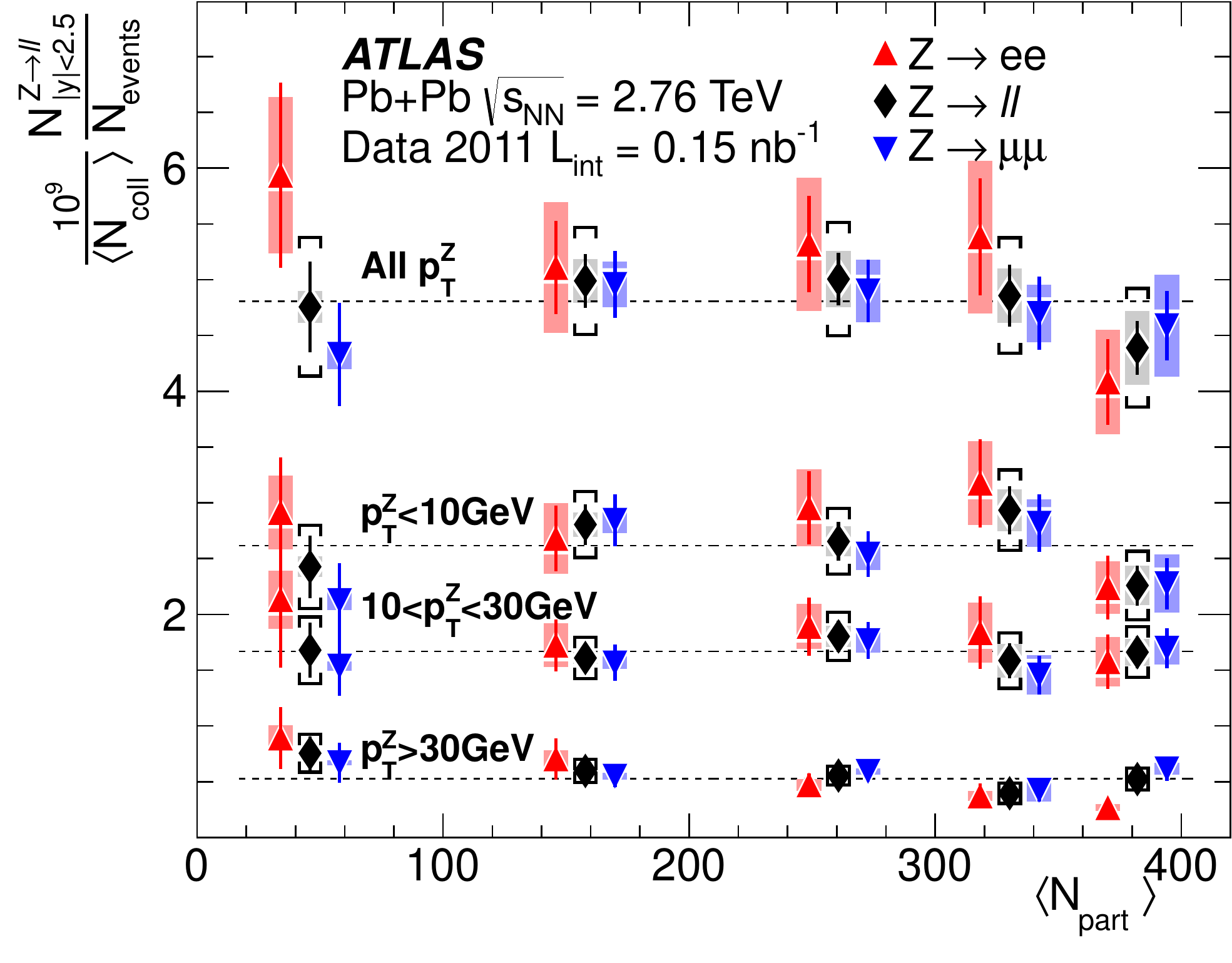}
\caption{Centrality dependence of $Z$ boson yields divided by \Ncoll. Results for $ee$ (upward pointing triangles) and $\mu\mu$ (downward pointing triangles) channels are shifted left and right, respectively, from their weighted average (diamonds).  
Bars and boxes represent statistical and systematic uncertainties, respectively.  For the combined results,  the brackets show the combined uncertainty including the uncertainty on \Ncoll, and the dashed lines show the results of fits, using a constant.} 
\label{fig:scaling}
\end{center}
\end{figure}

To further examine the binary collision scaling of the data, the $Z$ boson per-event yields, divided by \Ncoll, are shown in Fig.~\ref{fig:scaling} as a function of \Npart, in several \ptZ\ bins.
The figure demonstrates that the \zee\ and \zmm\ results are consistent within their uncertainties for all \ptZ\ and centrality regions.   
Within the statistical significance of the data sample, the $Z$ boson per-event yield obeys binary collision scaling. 
 
The elliptic anisotropy, $v_2$, of the $Z$ boson is defined as $v_{2} = \langle \cos 2(\phi - \Psi_2)\rangle/ \sigma_{2}$, where $\phi$ is the azimuthal angle of the $Z$ boson momentum vector and $\Psi_2$ is the azimuthal angle of the event plane, the plane containing the momentum vectors of both lead nuclei and measured with resolution $\sigma_{2}$ \cite{v2published}.  The $v_2$ values measured in the two decay channels are consistent and are combined. 
The main uncertainty on the $v_2$ measurement  arises from the event plane (EP) resolution,
which is measured from the difference of $\Psi_2$ determined using the two sides of the FCal at positive and negative rapidities \cite{v2published}.  To ensure that the jets associated with $Z$ boson production do not affect the determination of $\Psi_2$, the EP resolution is also measured ordering the FCal sides in the difference according to the direction of the $Z$ boson. A systematic uncertainty of 12~mrad is assigned for possible EP distortion.

\begin{figure}[ht]
\begin{center}
\includegraphics[width=0.49\textwidth]{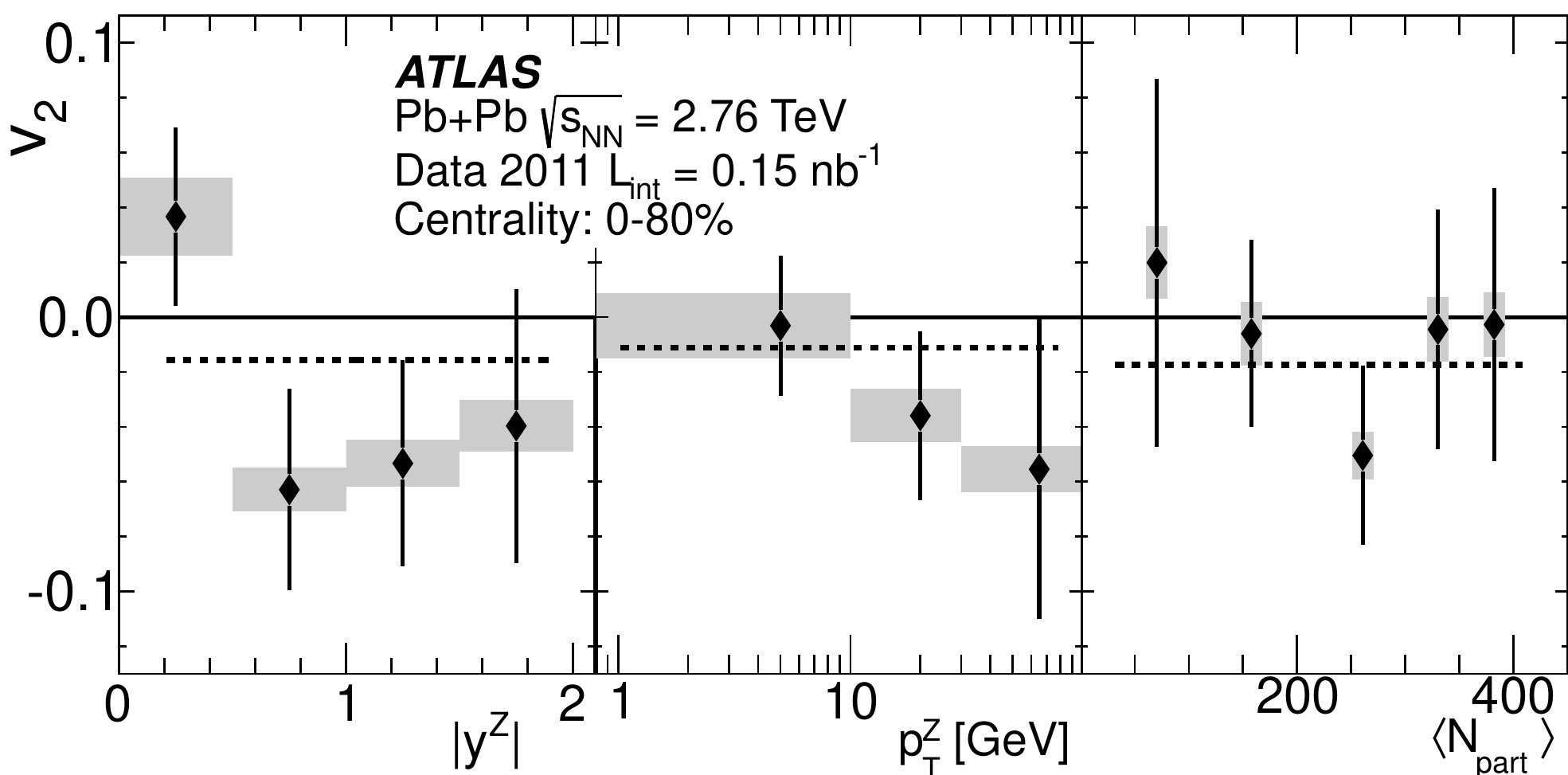}
\caption{$v_2$ as a function of $|\yZ|$ (left), \ptZ\ (center), and \Npart\ (right). Bars and boxes represent statistical and systematic uncertainties, respectively. The dashed lines show the results of constant fits to the $v_2$ values, considering only statistical uncertainties.}
\label{fig:flow}
\end{center}
\end{figure}

The $v_2$ of the $Z$ boson is shown in Fig.~\ref{fig:flow} as a function of $|\yZ|$, \ptZ, and \Npart. 
The averaged $v_2$ of the $Z$ boson 
has been measured to be $v_2=-0.015 \pm 0.018$(stat.)$\pm 0.014$(sys.), which indicates an isotropic distribution.
This observation is an independent measurement consistent with \zll\ yields being unaffected by the medium in HI collisions.

Using the ATLAS detector, $Z$ boson production has been measured in \PbPb\ collisions at \sqn=2.76\,\TeV\ using 0.15\,\inb\ of integrated luminosity collected in the 2011 LHC physics run. Within $|\yZ|<2.5$, and $66<m_{\ell\ell}<102$\,\GeV, a total of 772 and 1223 $Z$ boson candidates are reconstructed in the \zee\ and \zmm\ channels, respectively. The combinatorial background is at the level of 5\% in the  di-electron channel and 1\% for the di-muon channel. The $Z$ boson production yield integrated over $|y^{Z}|<2.5$ is consistent between the two channels in all measured \pt\ and centrality regions. The momentum and rapidity distributions of the $Z$ bosons 
are consistent with {\sc Pythia} simulations of $Z$ boson production in \pp\ collisions scaled to the NNLO cross section and multiplied by \mTAA. Within the uncertainties the $Z$ boson yield is found to be proportional to \Ncoll. The elliptic anisotropy of the $Z$ boson measured as a function of rapidity, \ptZ\ and \Npart\ is consistent with zero within the uncertainties of the measurements.



We thank CERN for the very successful operation of the LHC, as well as the
support staff from our institutions without whom ATLAS could not be
operated efficiently.

We acknowledge the support of ANPCyT, Argentina; YerPhI, Armenia; ARC,
Australia; BMWF and FWF, Austria; ANAS, Azerbaijan; SSTC, Belarus; CNPq and FAPESP,
Brazil; NSERC, NRC and CFI, Canada; CERN; CONICYT, Chile; CAS, MOST and NSFC,
China; COLCIENCIAS, Colombia; MSMT CR, MPO CR and VSC CR, Czech Republic;
DNRF, DNSRC and Lundbeck Foundation, Denmark; EPLANET and ERC, European Union;
IN2P3-CNRS, CEA-DSM/IRFU, France; GNSF, Georgia; BMBF, DFG, HGF, MPG and AvH
Foundation, Germany; GSRT, Greece; ISF, MINERVA, GIF, DIP and Benoziyo Center,
Israel; INFN, Italy; MEXT and JSPS, Japan; CNRST, Morocco; FOM and NWO,
Netherlands; BRF and RCN, Norway; MNiSW, Poland; GRICES and FCT, Portugal; MERYS
(MECTS), Romania; MES of Russia and ROSATOM, Russian Federation; JINR; MSTD,
Serbia; MSSR, Slovakia; ARRS and MVZT, Slovenia; DST/NRF, South Africa;
MICINN, Spain; SRC and Wallenberg Foundation, Sweden; SER, SNSF and Cantons of
Bern and Geneva, Switzerland; NSC, Taiwan; TAEK, Turkey; STFC, the Royal
Society and Leverhulme Trust, United Kingdom; DOE and NSF, United States of
America.

The crucial computing support from all WLCG partners is acknowledged
gratefully, in particular from CERN and the ATLAS Tier-1 facilities at
TRIUMF (Canada), NDGF (Denmark, Norway, Sweden), CC-IN2P3 (France),
KIT/GridKA (Germany), INFN-CNAF (Italy), NL-T1 (Netherlands), PIC (Spain),
ASGC (Taiwan), RAL (UK) and BNL (USA) and in the Tier-2 facilities
worldwide.

\bibliographystyle{atlasBibStyleWoTitle}
\bibliography{bibliography}

\begin{thebibliography}{10}
\bibitem{ATLAS-coord}
 The ATLAS reference system is a Cartesian right-handed coordinate system, with
  the nominal collision point at the origin. The anticlockwise beam direction
  defines the positive $z$-axis, while the positive $x$-axis is defined as
  pointing from the collision point to the center of the LHC ring and the
  positive $y$-axis points upwards. Transverse quantities, such as
  $p_{\mathrm{T}}$ and $E_{\mathrm{T}}$, are defined in the ($x$,$y$) plane.
  The azimuthal angle $\phi$ is measured around the beam axis, and the polar
  angle $\theta$ is measured with respect to the $z$-axis. The rapidity is
  given by $y = \frac{1}{2}ln\frac{E+p_z}{E-p_z}$ and pseudorapidity is defined
  as $\eta = - \ln(\tan(\theta/2))$.
\end{thebibliography}

\onecolumngrid
\clearpage
\begin{flushleft}
{\Large The ATLAS Collaboration}

\bigskip

G.~Aad$^{\rm 48}$,
T.~Abajyan$^{\rm 21}$,
B.~Abbott$^{\rm 111}$,
J.~Abdallah$^{\rm 12}$,
S.~Abdel~Khalek$^{\rm 115}$,
A.A.~Abdelalim$^{\rm 49}$,
O.~Abdinov$^{\rm 11}$,
R.~Aben$^{\rm 105}$,
B.~Abi$^{\rm 112}$,
M.~Abolins$^{\rm 88}$
% ...
\bigskip

% ...
$^{11}$ Institute of Physics, Azerbaijan Academy of Sciences, Baku, Azerbaijan\\
$^{12}$ Institut de F\'{i}sica d'Altes Energies and Departament de F\'{i}sica de la Universitat Aut\`{o}noma de Barcelona and ICREA, Barcelona, Spain\\
% ...
$^{21}$ Physikalisches Institut, University of Bonn, Bonn, Germany\\
% ...
$^{48}$ Fakult\"{a}t f\"{u}r Mathematik und Physik, Albert-Ludwigs-Universit\"{a}t, Freiburg, Germany\\
$^{49}$ Section de Physique, Universit\'{e} de Gen\`{e}ve, Geneva, Switzerland\\
% ...
$^{88}$ Department of Physics and Astronomy, Michigan State University, East Lansing MI, United States of America\\
% ...
$^{105}$ Nikhef National Institute for Subatomic Physics and University of Amsterdam, Amsterdam, Netherlands\\
% ...
$^{111}$ Homer L. Dodge Department of Physics and Astronomy, University of Oklahoma, Norman OK, United States of America\\
$^{112}$ Department of Physics, Oklahoma State University, Stillwater OK, United States of America\\
% ...
$^{115}$ LAL, Universit\'{e} Paris-Sud and CNRS/IN2P3, Orsay, France\\
% ...
\end{flushleft}


\end{document}